\documentclass[onecolumn]{aastex631}

\usepackage[section]{placeins}
\usepackage{upgreek}
\usepackage{CJK}
\usepackage{epstopdf}
\usepackage{amsmath}
\usepackage[version=4]{mhchem}
\usepackage[T1]{fontenc}

\begin{document}
\begin{CJK*}{UTF8}{gbsn}

\title{Feeling the Pressure: Effects of Formation Pressure on the Physical Properties of Titan Haze Analogs}

\correspondingauthor{Xinting Yu}
\email{xinting.yu@utsa.edu}

\author[0009-0003-9055-7397]{Adis Husi\'c}
\affiliation{Department of Physics and Astronomy, University of Texas at San Antonio\\
1 UTSA Circle, San Antonio, TX, 78249, USA}

\author[0000-0002-7479-1437]{Xinting Yu (余馨婷)}
\affiliation{Department of Physics and Astronomy, University of Texas at San Antonio\\
1 UTSA Circle, San Antonio, TX, 78249, USA}

\author[0000-0001-8364-620X]{Ryan C. Blase}
\affiliation{Southwest Research Institute\\6220 Culebra Rd, San Antonio, TX, 78238, USA}

\author[0000-0002-3865-0274]{Edward L. Patrick}
\affiliation{Southwest Research Institute\\6220 Culebra Rd, San Antonio, TX, 78238, USA}

\author[0000-0002-9075-0723]{Eric Austin}
\affiliation{Department of Physics and Astronomy, University of Texas at San Antonio\\
1 UTSA Circle, San Antonio, TX, 78249, USA}

\author[0000-0003-2477-3043]{Alan G. Whittington}
\affiliation{Department of Earth and Planetary Sciences, University of Texas at San Antonio\\
1 UTSA Circle, San Antonio, TX, 78249, USA}

\begin{abstract}

The Cassini--Huygens mission detected large negative ions in Titan's ionosphere at pressures as low as $10^{-6}$~torr. These ions ultimately polymerize to form Titan's complex organic haze particles, which are observed throughout the atmosphere and potentially on the surface. Laboratory analogs of these hazes, known as \textit{tholins}, have been used to study Titan's aerosols; however, most are produced at much higher pressures. The influence of formation pressures on key physical properties---such as particle size, density, surface energy, and mechanical strength---remains poorly constrained. These properties govern the haze's aggregation efficiency, radiative behavior, and surface--atmosphere interactions, shaping Titan's climate and surface. To investigate the effects of formation pressure, we generate tholins using a newly developed cold plasma discharge system. A 95\% nitrogen and 5\% methane gas mixture is exposed to plasma at two pressures, 1~torr and 0.125~torr. For both samples, we measure the production rate, particle size, morphology, density, surface free energy, Young's modulus, and nanoindentation hardness. While particle size, morphology, surface energy, and Young's modulus are similar across both pressures, tholins produced at lower pressure exhibited a threefold lower production rate, but a higher density and nanoindentation hardness. These variations likely reflect pressure-dependent changes in chemical structure, porosity, and mechanical strength. Because Titan's hazes form at much lower pressures than investigated here, actual haze particles are potentially even denser and mechanically stronger than our analogs, with implications for aerosol aggregation, aeolian and fluvial transport, and surface modification on Titan.
\end{abstract}


\received{January 26, 2026}
\revised{April 15, 2026}
\accepted{April 21, 2026}
\submitjournal{Planetary Science Journal}

\section{Introduction}\label{sec:Intro}
Saturn's moon, Titan, is the only moon in the Solar System with a dense atmosphere. Ground based observations, flybys of Voyager 1 and 2, Pioneer 11, and the Cassini--Huygens joint orbiter-lander mission constrained the composition of the atmosphere to primarily molecular nitrogen (\ce{N2}, 94-98\%) and methane (\ce{CH4}, 1.8-6\%), with tertiary gases such as carbon monoxide (\ce{CO}, 50 ppm) and molecular hydrogen (\ce{H2}, 0.1-0.2\%) \citep{Kuiper44,Tomasko2005, Flasar2005, Strobel2010, Horst_2017}. This evolving view of Titan has revealed a chemical factory in Titan's atmosphere where solar ultraviolet irradiation, energetic particles trapped in Saturn's magnetic field, and cosmic rays all provide the energy to power chemical reactions between \ce{CH4} and \ce{N2} that form complex organic molecules (e.g., \cite{Nixon2024} and references within). 

The Cassini Plasma Spectrometer-Electron Spectrometer established a potential lower mass limit for negative ions of mass-to-charge ratio of $10,000$~amu/q at pressures as low as $10^{-6}$~Torr ($10^{-4}$~Pa), with an estimated size up to 100~nm \citep[][]{Coates2007, Coates2009}. Additional observations from Cassini's Ion and Neutral Mass Spectrometer (INMS) and Ultraviolet Imaging Spectrometer at these pressures also detected large organic aerosols in Titan's atmosphere {\citep[e.g.,][and references within]{Waite2007, Liang2007}. These low-pressure ions and aerosol} likely go on to form the hazes present all throughout Titan's atmosphere \citep[e.g.,][]{Sagan1993,Tomasko2005,Waite2007,Coates2009}.

Additionally, the Cassini mission mapped much of Titan's surface, revealing that up to 84\% contains some amount of organics \citep[][]{Lopes2020}. This includes plains \citep[${\sim}65$\% of Titan's total surface area,][]{Lopes2016}, equatorial dunes \citep[${\sim}17$\%,][]{Lorenz2006, Radebaugh2013}, and  so-called `labyrinth terrain' \citep[${\sim}1.5$\%,][]{Malaska2020}, all of which exhibit organic signatures. Lakes and seas of liquid methane-ethane were also discovered \citep[${\sim}1.5$\%,][]{Stofan2007}, primarily at Titan's poles. While it is currently not possible to determine the true composition or origin of these surface organics, the most likely source is from the deposition of the photochemical hazes that surround the moon \citep[e.g.,][]{Sagan1993,Tomasko2008, Lorenz2008}, allowing for interactions with Titan's geology and hydrology.

Despite the advances in our understanding of Titan over the last 80 years, there are still many open questions regarding the haze, surface organics, and their potential link. The upcoming Dragonfly mission \citep{Lorenz2018, Barnes2021} will attempt to resolve some of these questions by investigating the chemical composition of the surface organics. The two main mission targets are equatorial dunes in the Shangri-La region, composed of organics and potentially small amounts of water ice \citep{Soderblom2007, Janssen2009,Radebaugh2013, Bonnefoy2016}, and the Selk impact crater \citep{Lorenz2021}, a region where there is the potential to measure mixed organic material and water ice melt. 

While the Dragonfly mission will provide unprecedented \textit{in-situ} measurements of Titan's surface organics, significant limitations remain in fully characterizing their physical and chemical properties. On account of these difficulties, experiments have been conducted on Earth to create analogs of Titan's organics. The first of these was performed in the 1970s by Carl Sagan and Bishun Khare, who later named these analogs `tholins' \citep{SaganKhare1971, SaganKhare79}. Since then, many different setups, typically employing a plasma source to simulate the energetic particles and cosmic rays or an ultraviolet lamp source to simulate solar ultraviolet (UV) irradiation, have been constructed \citep[e.g.,][]{Clarke2000, Tran2003, Szopa2006, Sekine2008, Sciamma2010, HorstTolbert2014, Sciamma-O'Brien2014, He2017, SebreeC2018, Hong2018, Hirai2023, Ugelow2024, Yang2025}. More exotic energy sources such as gamma rays and laser-induced plasmas have been used previously \citep{Ramirez2001}, but the resulting products do not match as well as cold plasma sources to Huygens' Pyrolysis Gas Chromatograph-Mass Spectrometer measurements of Titan's haze \citep{Coll2013}.

Although Titan's hazes are produced in the upper atmosphere at much lower pressure (less than $10^{-6}$~Torr, $10^{-4}$~Pa), higher pressures are required in the laboratory to generate a sufficient amount of materials in a reasonable timeframe (i.e., weeks rather than months or years). This has led to many studies on tholins to employ pressures of 1~Torr or higher. Pressure can influence the collision frequency, reaction pathways, and ultimately the physical properties of the resulting tholins. Thus, in order to understand the effects of the formation pressure on the resulting material, we generated tholin samples at two pressures, 1~Torr (133.3~Pa), chosen since numerous previous studies investigating tholins have employed gas mixtures of this pressure \citep[e.g.,][]{Imanaka2004,Sciamma2010,He2017}, and 0.125~Torr (16.66~Pa), chosen to represent a factor of 10 lower pressure that would be more representative of the upper stratosphere nearer to Titan's detached haze layer \citep{Horst_2017}. 

We present the results of many physical properties of these two tholin sample sets, including production rate, particle size and morphology, density, surface free energy, Young's modulus, and nanoindentation hardness. These diagnostic properties will help us assess whether pressure plays an important role in shaping the physical properties of Titan's hazes. These properties have a direct impact on many important processes, such as saltation \citep{Lorenz2006,Kok2012}, haze-liquid interactions such as particle floatation \citep{Cordier2019, Yu2020}, and cloud condensation \citep{Lavvas2011,Yu2020,Li2022}. Therefore, understanding these fundamental properties of tholins may allow us to understand the role of organics in shaping the geological features of Titan.

This paper is structured as follows: We first describe our tholin synthesis and collection methods in Section \ref{sec:synthesis}. In Section \ref{sec:methods}, we describe the methods of scanning electron microscopy (Section \ref{sec:sem}), helium pycnometry (Section \ref{sec:pyc}), contact angle goniometry (Section \ref{sec:contact}), and thin film nanoindentation (Section \ref{sec:nanoindentation}). Section \ref{sec:results} covers the results for particle size and morphology (Section \ref{sec:psd}), density (Section \ref{sec:densityresult}), surface free energy and the partitioning components (Section \ref{sec:SurfaceEnergy}), and Young's modulus and nanoindentation hardness (Section \ref{sec:EandHResults}). Finally, in Section \ref{sec:discussion}, we discuss the differences observed between our high- and low-pressure samples (Section \ref{sec:hplpdiff}) and the implications for organic aerosols and its behavior on Titan (Section \ref{sec:implication}).

\section{Tholin Synthesis and Collection}\label{sec:synthesis}

\begin{figure}
\gridline{
\fig{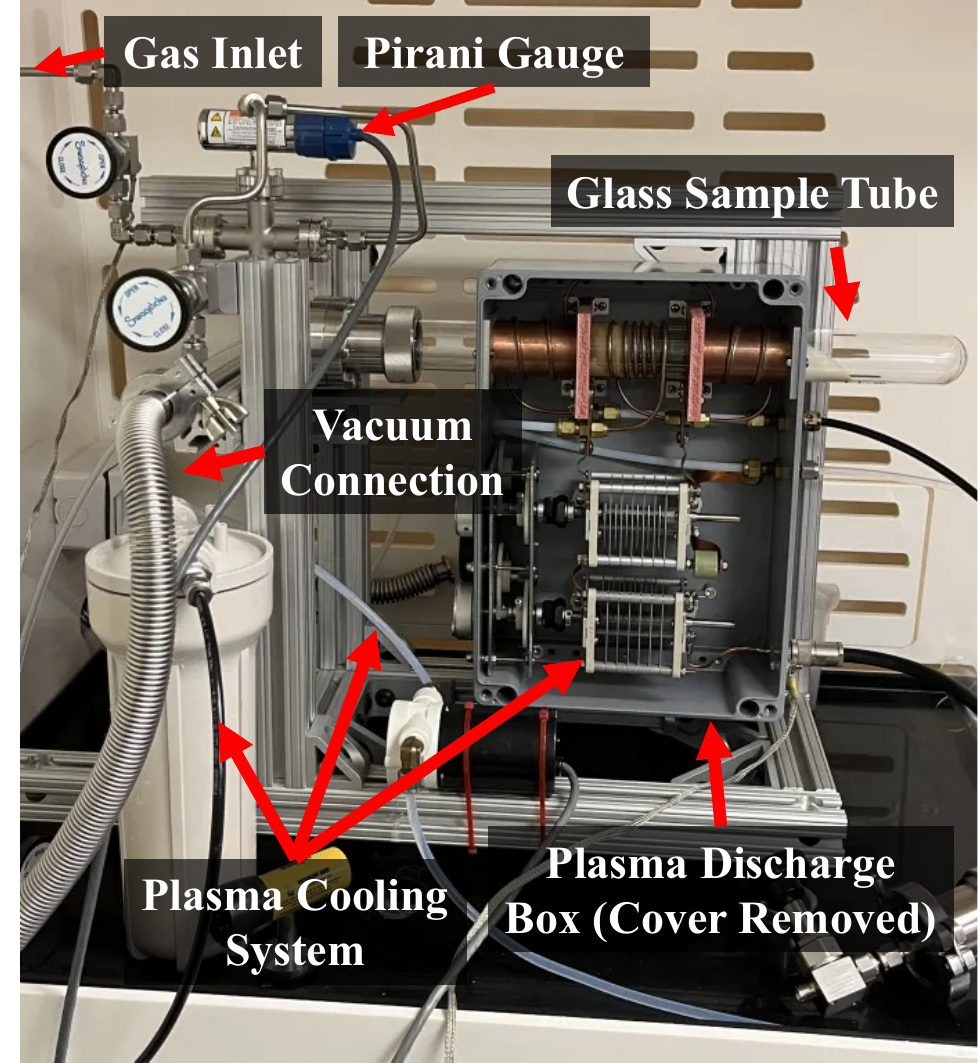}{0.6\textwidth}{(a)}}
\gridline{
\fig{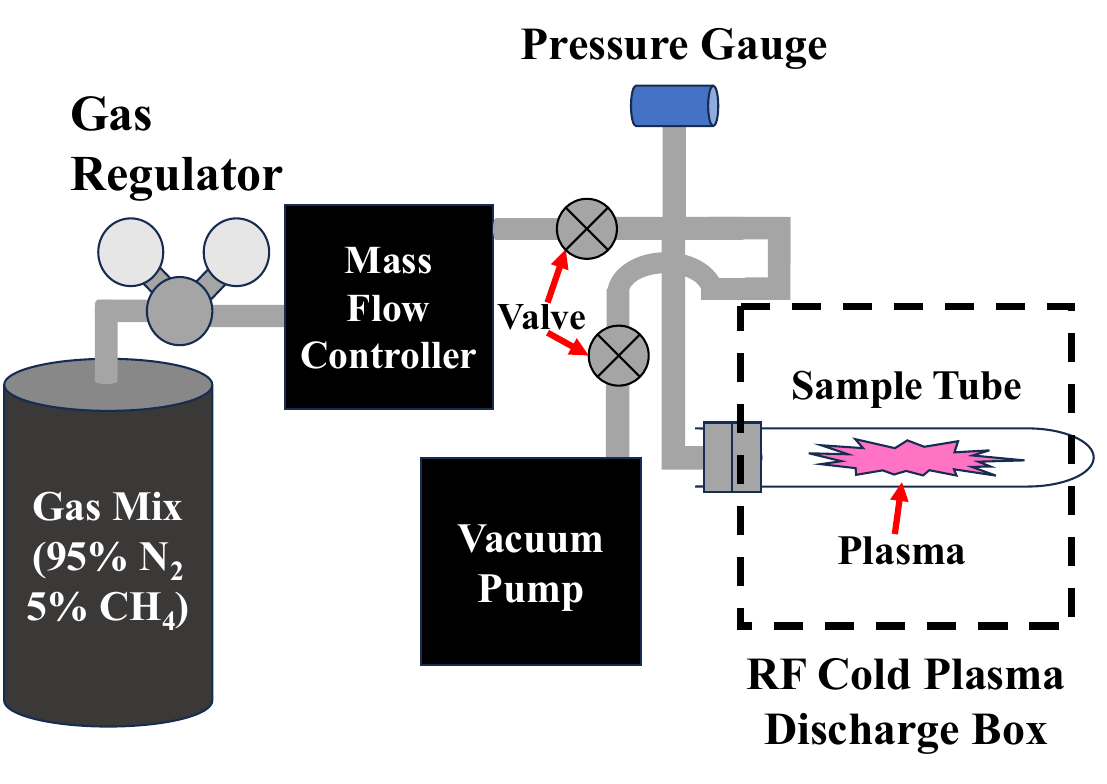}{0.6\textwidth}{(b)}}
\caption{\textbf{(a)} Annotated Tholinator setup at the Southwest Research Institute. Not pictured are the supply gas cylinder and mass flow controller (out of frame, to the left) and the vacuum pump (out of frame, below). The plasma discharge box typically has a cover in place to protect the system components and users. \textbf{(b)} Simplified schematic diagram of the Tholinator setup.}
\label{fig:Tholinator}
\end{figure}

In this work, we utilized the previously unpublished Tholinator experimental setup, housed at the Southwest Research Institute's Space Environment Simulation Lab (SwRI SESL). Figure \ref{fig:Tholinator}a shows an image of the experimental setup. Figure \ref{fig:Tholinator}b shows the schematic diagram of the system. The energy source is a cold radio frequency (RF) plasma driven by an Advanced Energy RFX600-A RF supply. The RF frequency is $13.56\pm0.05\%$~MHz and can provide up to 600~W of forward power. For these experiments, we used a forward power of 36~W to maximize the available power without causing any arc discharges within the tube. Some studies suggest that cold RF plasmas and $\ce{N2}$/$\ce{CH4}$ gas mixtures produce non-negligible amounts of UV photons that contribute to the photochemistry \citep[e.g.,][]{Clay1996, Imanaka2004}. Therefore, it is possible that the setup emulates a wide variety of the constant energy source chemistry on Titan but excludes sporadic high-energy events such as lightning or meteor strikes \citep{Cable2012}.

Previous studies show that cold plasma-generated tholins better replicate results from the Huygens Aerosol Collector Pyrolyser and Gas Chromatograph-Mass Spectrometer \citep{Coll2013}. In addition, tholins produced with mixtures of 1\% and 5\% \ce{CH4} concentration have similar ion chemistry to those measured by Cassini's INMS instrument \citep{Dubois2020}. Therefore, cold plasma energy sources such as the one used in the Tholinator may create more authentic analogs when attempting to simulate Titan's bulk atmospheric chemistry. Other sources such as UV lamp irradiation may be better at recreating chemistry that occurs deeper in Titan's atmosphere due to UV radiation's ability to penetrate into regions as low as Titan's stratosphere \citep{Horst_2017}.

The Tholinator is a flow-type system where new gas is continuously introduced to the system, exposed to the plasma, and then pumped out by a vacuum pump. Tholin synthesis takes place within a quartz glass tube that is open on one end with connections to a convection-enhanced Pirani pressure gauge (MKS Granville Phillips 275 Convectron Gauge), vacuum pump (Anesta Iwata ISP 250C Scroll Pump), and gas inlet, while the other end is closed. The tube dimensions are 430~mm in length with an inner diameter of 35~mm, and an outer diameter of 38~mm. 

Gas inflow is controlled using a Sierra SmartTrak 100 digital mass flow controller with $\pm 0.5$\% accuracy. The Tholinator may use any of the commonly used mixtures for Titan \citep[see][for an overview]{Yu2023}, but for this study, we focus on tholins synthesized using a 95\% nitrogen (\ce{N2}) and 5\% methane (\ce{CH4}) (99\% purity $\pm$ 0.1\%, Airgas) mixture. The flow rate of the gas is adjusted using the mass flow controller and vacuum pump in order to keep the pressure within the glass tube at either approximately 1~Torr (133.3~Pa), referred to as `high-pressure' or `HP', or approximately 125~mTorr (16.67~Pa), referred to as `low-pressure' or `LP'. For high-pressure experiments, the flow rate is 7 standard cubic centimeters per minute (sccm, where 1~sccm is 1~cm$^3\cdot$min$^{-1}$ at $0$~$^\circ$C and 1~atm), while for low-pressure experiments the flow rate is 2~sccm.

Residence time is related to pressure, and can be estimated by the following equation from \citet{He2022}:
\begin{equation}\label{eq:residencetime}
    t=\frac{VP_1}{QP_0}
\end{equation}
where $V$ is the volume of the chamber, $P_0$ is the standard pressure (760~Torr), $P_1$ is the pressure in the chamber, and $Q$ is the flow rate (typically in sccm). Using Equation \ref{eq:residencetime}, the Tholinator has a gas residence time of approximately 47~s at 1~Torr and 16~s at 0.125~Torr.

Tholin synthesis takes place inside the glass tube as long as the plasma is maintained. Typical experiment run times of the low-pressure case are approximately 500 hours, while for the high-pressure case they are approximately 170 hours. These times are chosen to produce approximately 1~g of material, with low-pressure experiments having lower production rates of approximately 2~mg$\cdot$hr$^{-1}$ and high-pressure experiments having a rate of approximately 6~mg$\cdot$hr$^{-1}$. A comparison of the Tholinators production rates to previously published rates is provided in Table \ref{tab:prodrates}. Our high-pressure production rate is similar to those of previously published experimental setups that utilize a plasma energy source \citep[e.g.,][]{He2017}. 

Titan itself has an estimated aerosol production rate of $9\times10^{10}$~mg$\cdot$hr$^{-1}$, calculated by using the reported column averaged production rate of aerosols from \citet{Lavvas2010}, $3\times10^{-14}$~$\text{ g cm}^{-2}\text{ s}^{-1}$, multiplied by Titan's surface area, $8.33\times10^7$~cm$^2$. The production rate of aerosols on Titan, when limited to a surface area similar to the Tholinator reaction vessel, $513$~mm$^2$, would be ${\sim} 5.5\times10^{-7}$~mg$\cdot$hr$^{-1}$, meaning the Tholinator produces over 3,000,000 to 10,000,000 times more material than is produced in Titan's atmosphere. In other words, 1 second in the Tholinator is equivalent to approximately 800 to 2800 hours on Titan.
\begin{deluxetable}{lcccr}[h]
    \tablewidth{0pt}
    \tablecaption{Production rate comparisons between the Tholinator (this work) and other cold RF plasma experimental setups, where available. \label{tab:prodrates}}
    \tablehead{
    \colhead{Setup} & \colhead{Production Rate} & \colhead{Gas Mixture} & \colhead{Pressure} & \colhead{Source} \\
    \colhead{} & \colhead{(mg$\cdot$hr$^{-1}$)} & \colhead{(\ce{N2}/\ce{CH4})} & \colhead{(Torr)} & \colhead{}}
    \startdata
        Tholinator  & 2 & 95/5  & 0.1 &This work\\
        Tholinator &$6$ & 95/5 & 1 &This work\\
        PHAZER &$7.25 - 7.42$& 95/5 & 2 & \citet{He2017}\\
        PAMPRE\tablenotemark{a} &$28.2$& 94/6 & 0.68 & \citet{Sciamma2010}\\
        PAMPRE\tablenotemark{a} & $49.3$ & 94/6 & 1.28 & \citet{Sciamma2010}\\
    \enddata
    \tablenotetext{a}{Reported rate at most similar gas mixture values and pressures to the Tholinator.}
\end{deluxetable}

During the experiment, synthesized tholin is deposited on the inner walls of the glass tube. Within the synthesis tube we may place a number of pre-prepared substrates, typically glass slides or silicon wafers, to allow for thin film deposition. Once synthesis is complete, the glass tube may be removed from the Tholinator apparatus. Any substrates present are collected and then placed within a vacuum desiccator to limit atmospheric water and oxygen contamination. A metal endcap is then attached to the synthesis tube's open end and the tube is kept under vacuum to prevent any contamination from atmospheric water and oxygen.

\begin{figure}[htb]
    \centering
    \gridline{
    \fig{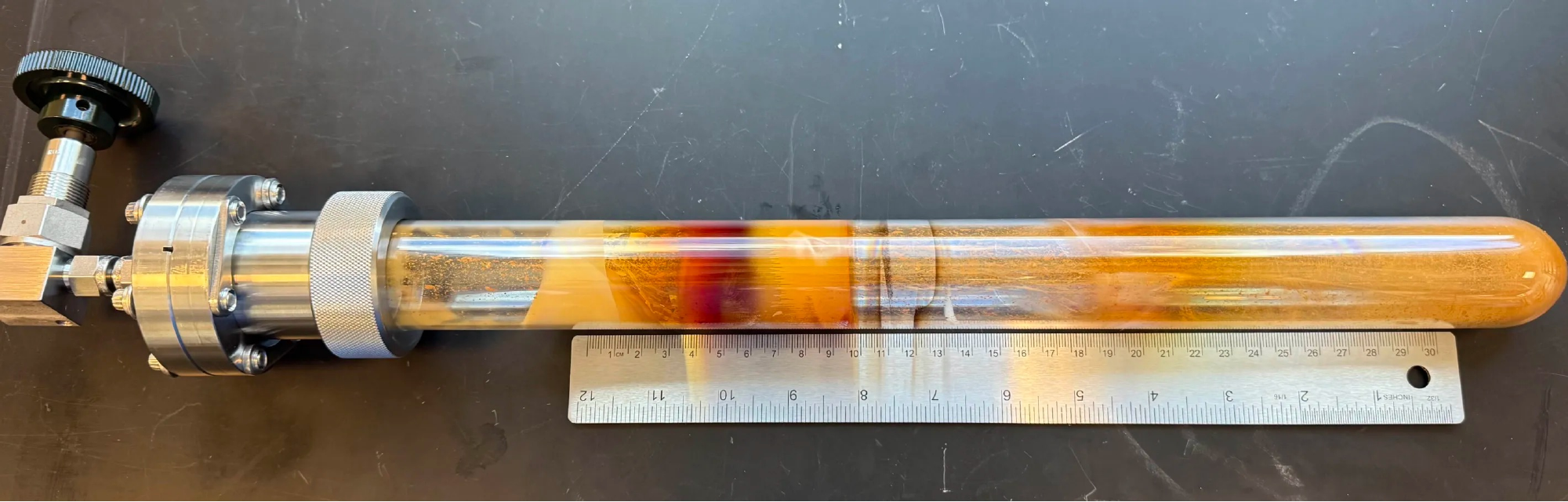}{0.92\textwidth}{(a)}
    }
    \gridline{
    \fig{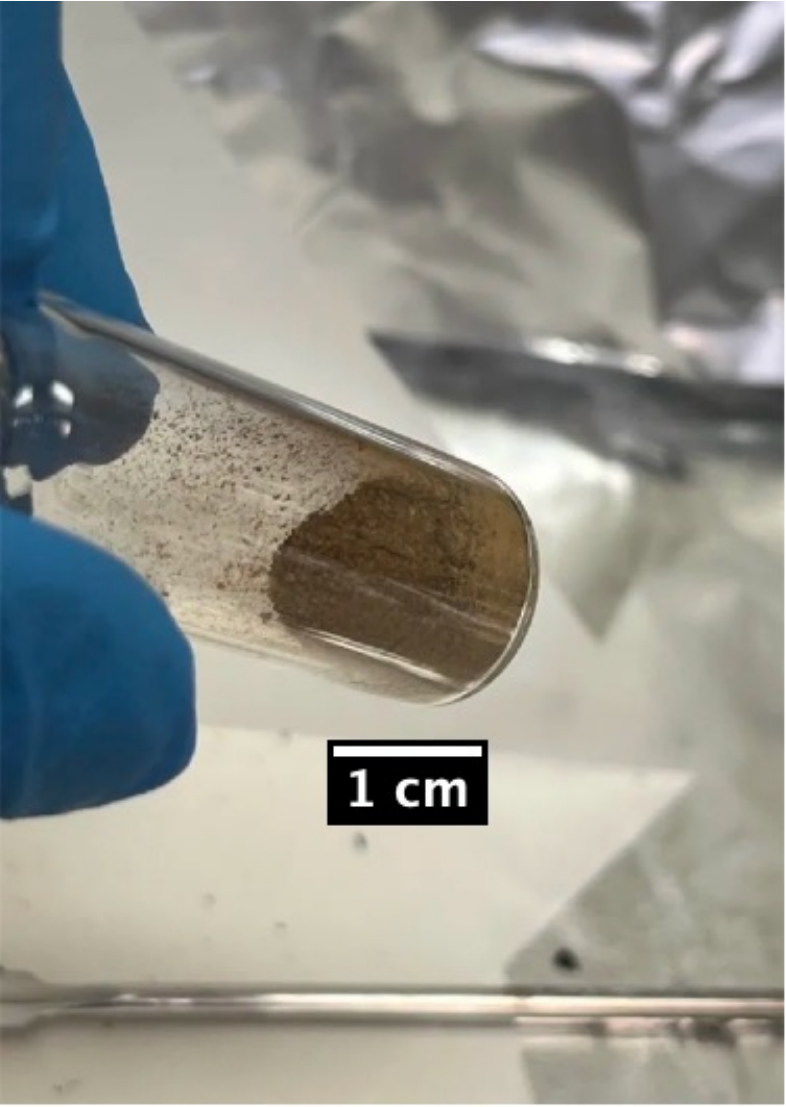}{0.42\textwidth}{(b)}
    \fig{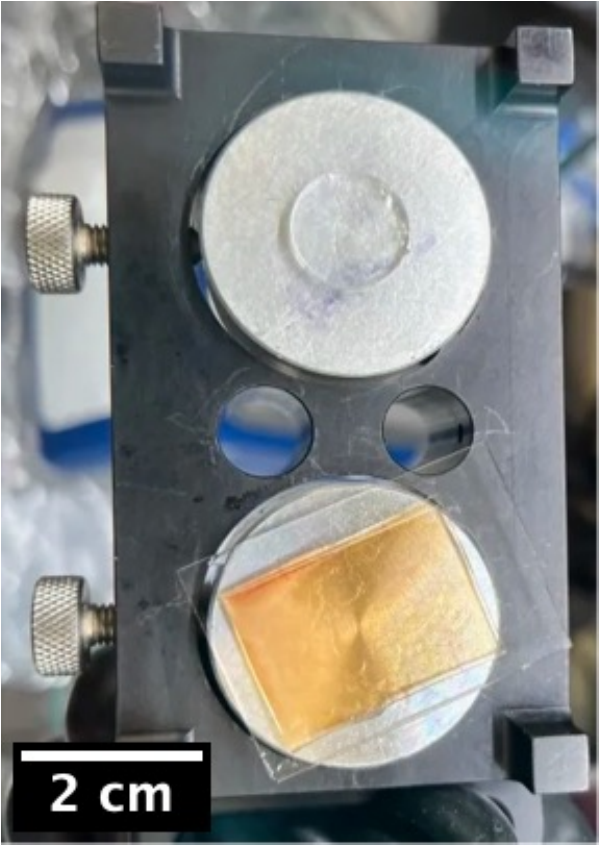}{0.42\textwidth}{(c)}
    }
    \caption{\textbf{(a)} A tube with synthesized high-pressure material (tan and red regions) before collection. The origin of the color difference is unclear, but it is thought to be due to a difference of the thickness/amount of deposited material in each region. Black markings (around 10.7 and 12~cm) are from highly localized arc discharging where the RF plasma electrodes are located, and only affect the glass tube itself. Significant arcing has not been witnessed and is not expected to occur or affect resulting tholin. The run time of this experiment is 180~hours. \textbf{(b)} Powder samples obtained after manual scraping of the red and tan regions in subfigure a. Scraped powders achieve a uniform color with no clear tan or red differences after collection. \textbf{(c)} A fused silica standard (top) and tholin thin film on glass (bottom), ready for nanoindentation measurements.}
    \label{fig:tubePowderAndFilm}
\end{figure}

Powder samples are manually collected at a later time by physically scraping the synthesis tube walls using a cleaned lab spatula. We first sonicate the tube with the endcap attached, still under vacuum, in order to loosen as much material as possible and minimize manual collection time. Previous work has shown that tholin samples exposed to air oxidize slowly but rapidly adsorb water, and that this adsorbed water can be removed through vacuum pumping for a minimum of 30 minutes \citep{chatain}. Therefore, for samples that are utilized in measurement methods that are highly sensitive to water adsorption, such as surface free energy, we collected the sample in an inert environment \ce{N2} gas glovebox (<0.1\%~relative humidity of \ce{H2O}), and then stored it under vacuum. Samples used in experiments that would result in atmospheric exposure regardless, such as density and particle size measurements, are collected in Earth's atmosphere with a minimized collection time (<1~hour) and then stored under vacuum for a minimum of 24 hours. Figure \ref{fig:tubePowderAndFilm} shows examples of a tube with synthesized material, collected powders, and a film on a glass substrate.

\begin{figure}
    \centering
    \includegraphics[width=1\linewidth]{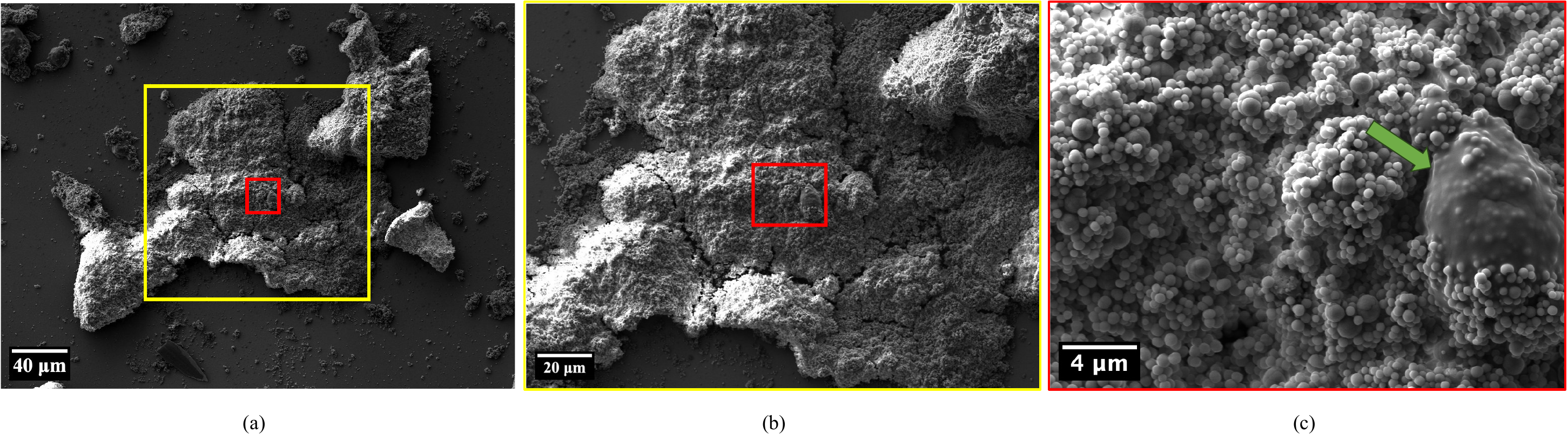}
    \caption{\textbf{(a)} An example image of a large wall particle scraped from the wall of the deposition tube. These particles are large and flat, providing sites for smaller 1~$\upmu$m particles to grow from or attach to. The yellow inset box is subfigure (b), and the red inset box is subfigure (c). \textbf{(b)} A more magnified image of the wall particle showing smaller, 1 micron spherical particles attached to the wall particle. The red inset box is subfigure (c). \textbf{(c)} A more magnified image of the wall particle, with identifiable spherical particles attached. The green arrow indicates the exposed, smooth area of the wall particle.}
    \label{fig:wallparticle}
\end{figure}

We note here that tholin samples come in two distinct forms: spherical particles and jagged, large particles called `wall particles' (see Sections \ref{sec:sem} and \ref{sec:psd} for a more thorough discussion on morphology). Figure \ref{fig:wallparticle} shows an example of one such wall particle. Although spherical particles tend to naturally form aggregates of multiple individual spherical particles, wall particles are thought to be created as a result of the confining chamber walls and have been noted in other experimental setups \citep[e.g.,][]{Vanssay1999, Szopa2006, Nguyen2007, Cable2012}. Their size is significantly larger than that of individual spherical aggregates. Therefore, collected powders may either be spherical particles or, if collected by scraping, may include a fraction of wall particles. On the other hand, thin films are thought to be purely composed of wall particles, with the uppermost layer being a formation site for spherical particles.

\section{Experimental Methodologies}\label{sec:methods}
In this section, we describe the experimental techniques used to characterize synthesized tholin films and powders. These methods include scanning electron microscopy for particle size and morphology (Section \ref{sec:sem}), helium pycnometry for density (Section \ref{sec:pyc}), contact angle goniometry for surface free energy (Section \ref{sec:contact}), and nanoindentation for Young's modulus and nanoindentation hardness (Section \ref{sec:nanoindentation}).

\subsection{Particle Morphology and Size Distribution}\label{sec:sem}
In the following section, we define `particles' as the spheroidal objects present in scanning electron microscopy images. `Monomers' are defined as the objects that particles are composed of. `Aggregates' are then defined as collections of bound particles. An example of an aggregate with multiple particles can be found in Figure \ref{fig:particlesizes}a, and Figure \ref{fig:particlesizes}b shows an example of one particle with visible monomers comprising the surface (one such monomer is outlined in red).

Individual tholin particles for both low- and high-pressure samples were imaged using a Zeiss Crossbeam 340 Scanning Electron Microscope (SEM) at the University of Texas at San Antonio's Kleberg Advanced Microscopy Center. Powder samples were placed onto carbon tape and coated with either ${\sim}20$~nm of carbon using an SPI Module Carbon Coater or ${\sim}10$~nm of gold using a Pelco SC-7 Sputter Coater to reduce charging effects. Gold coating was used when elemental analysis was employed or when high magnification imaging was necessary, such as for imaging monomers. Some charge artifacts are still present in some SEM images due to height differences or uneven coating of the sample.

Particle diameters were manually measured using the FIJI distribution of ImageJ \citep{Schneider2012, Schindelin2012} to create a particle size distribution histogram for high- and low-pressure samples and a mean and median diameter were calculated. Over 1000 particles were measured in order to obtain a statistically significant result. Monomer diameters were obtained from samples that were gold-coated.

A standard procedure was followed for counting particle sizes. To avoid selection bias, if an image was of good quality (i.e., particles were distinguishable from the background and any contaminating particles), then every particle that was visible was counted by fitting a perfect circle around the particle and then measuring the diameter of the fitted circle. The diameters reported here are the equivalent circular diameters. Due to a consistent spherical shape among particles, this method allows the measurement of particles that are either slightly malformed or somewhat obscured by other particles. For slightly obscured particles, we use the equivalent circle which fits the visible edges of the particle. Particles that are cleaved in half or otherwise significantly malformed or obscured are not counted here. An example of this measurement technique is given in Figure \ref{fig:particlesizes}a.

\subsection{Density}\label{sec:pyc}
The density of collected powder samples is measured using an Anton Paar Ultrapyc 3000 helium pycnometer. Due to constraints in the amount of tholin material available, we use the nano cell (0.25~cm$^3$). The nano cell is calibrated with a standard steel calibration ball with a known volume, provided by Anton Paar. 

To measure powdered tholin samples, we first measure the mass of the empty nano cell and then the mass of the nano cell when it is filled to approximately 70\% in volume with tholin. The mass of our sample, $m$, is then obtained by taking the difference of these two measurements. We then performed multiple helium gas fill and purge cycles on our sample to remove any adsorbed water. Once a stable volume measurement was obtained, we carried out the pycnometry experiments to measure the sample volume. After the pycnometry experiments were completed, the mass of the cell was measured again to quantify the change in mass due to water loss, if any.

A minimum of 15 measurements were completed using the pycnometer to obtain the sample volume and take the weighted average, $\bar{V}$, of these measurements. The density of the sample is then given by $\rho = m/\bar{V}$. The resulting powder density is the skeletal density, measuring the volume of the particle excluding any internal voids or pores that are not accessible to the helium gas.

\subsection{Surface Free Energy}\label{sec:contact}
Surface free energy, $\gamma$, is defined as the change in free energy ($W$) when the surface area of a solid ($A$) is increased by a unit area, $\gamma = dW/dA$. This change is equivalent to the energy needed to separate two contacting solid surfaces per two unit areas, which can then be used to determine adhesion and wettability of the film. Surface free energy has previously been measured for other tholin samples \citep{Yu2017,Yu2020,Li2022}. 

We performed contact angle goniometry by employing the sessile drop method using an Ossila Contact Angle Goniometer to measure contact angles between liquids and thin tholin films deposited on substrates. One polar and one nonpolar liquid can be used for this method to provide constraints on the polar and dispersive components of the sample. By employing the Owens-Wendt-Rabel-Kaelble (OWRK) method \citep{OwensWendt1969,Kaelble_1970,rabel1971}, we may then calculate the polar and dispersive components of the surface free energy, $\gamma_p$ and $\gamma_d$, using the measured contact angles. A more rigorous treatment of the analytical solutions to the OWRK method is given in the appendix of \citet{Li2022}.

We first set up the experimental equipment in a temporary glovebox filled with ultra high purity dry \ce{N2} (99.999\% purity, Airgas) providing an inert environment with a relative humidity of less than 0.1\%, monitored with a hygrometer. This creates an environment in which water adsorption and sample contamination can be minimized. Contact angle measurements were then performed using the goniometer in this environment. 

We used the nonpolar liquid diiodomethane (\ce{CH2I2}; 99\% purity, Fisher Chemical) and polar liquid water (HPLC Grade Water, Fisher Chemical) when performing these experiments. Videos of the experiment were taken at 30 frames per second in order to analyze the initial contact angle and its evolution over time. At the start of the video, a single sessile droplet is dispensed using a pipette. Ossila Contact Angle Software 4.0 is then used to analyze each frame of the surface contact angle measurement. The contact angle in each frame is averaged together to obtain a single average contact angle value for the drop. A minimum of three drops of each liquid are averaged together to provide the final contact angle result. The diiodomethane drops were recorded for 60 seconds, while the HPLC water drops were recorded for 30 seconds. The diiodomethane droplets were analyzed over the full duration of the video due to minimal evaporation and negligible dissolution of the tholin sample. In contrast, water droplets were analyzed only during the initial 2.5~s to minimize the effects of tholin dissolution and droplet evaporation \citep{Li2022}.  Figure \ref{fig:SurfaceEnergies}a shows an example of a singular drop and corresponding angle measurements. 

\subsection{Young's Modulus and Nanoindentation Hardness}\label{sec:nanoindentation}
Young's modulus, $E$, is the measure of a material's response to being stretched or compressed by a force in one axis. Formally, Young's modulus is the ratio of the tensile or compressive stress, $\sigma$, over the axial strain, $\epsilon$,
\begin{equation}
    E=\frac{\sigma}{\epsilon}
\end{equation} 
A low Young's modulus material is typically able to stretch or compress easily (e.g., rubber). Conversely, high Young's modulus materials are more resistant to stretching or compression (e.g., steel). Nanoindentation hardness, $H$, is a materials resistance to plastic (i.e., `permanent') deformation or abrasion. Low nanoindentation hardness materials include rubber, talc, and gypsum, while high nanoindentation hardness materials include steel, silica, and quartz. Nanoindentation hardness and Young's modulus are often positively correlated, meaning materials with higher nanoindentation hardness typically have higher Young's modulus \citep[See, for example, ][]{Yu2018, Yan2022}. 

To measure these properties, we employed nanoindentation. Nanoindentation has previously been performed on tholin, simple organics, and common Earth sands in \citet{Yu2018}, and the general methods are described in \citet{OliverPharr1992, OliverPharr2004}. Here we use similar methodologies and sample preparation techniques, using a KLA Instruments iMicro nanoindenter. 

In order to perform nanoindentation, we make use of a thin tholin film deposited onto a substrate. For thin films on substrates, it is possible to remove much of the substrate's influence on the thin film response when the Young's modulus and Poisson's ratio of the substrate are known \citep{HayCrawford2011}. Poisson's ratios of 0.28 for silicon wafers and 0.3 for glass slides were assumed. The Young's modulus of each substrate was measured via nanoindentation using the assumed Poisson's ratios.

We used a Berkovich diamond tip with a known geometry to create indentations into sample films. These indentations stop once either a specified load (50~mN) or specified depth into the sample (up to 40\% of the film's thickness) is reached. The resultant load-displacement curve can then be used to calculate nanoindentation hardness, $H$, and Young's modulus, $E$, of the sample. The details of this calculation are given in \citet{OliverPharr1992, OliverPharr2004, HayCrawford2011}, and \citet{Yu2018}.

Additionally, we used continuous stiffness measurements which perform a miniature loading-unloading cycle at a rate of 5~Hz during the entire measurement, providing nanoindentation hardness and modulus values as a function of depth \citep{OliverPharr2004}. We performed indentation at a minimum of 70 points across each sample and calculated a weighted average to obtain a statistically significant result.

\section{Results}\label{sec:results}
\subsection{Particle Morphology and Size Distribution}\label{sec:psd}

\begin{figure}
\gridline{
\fig{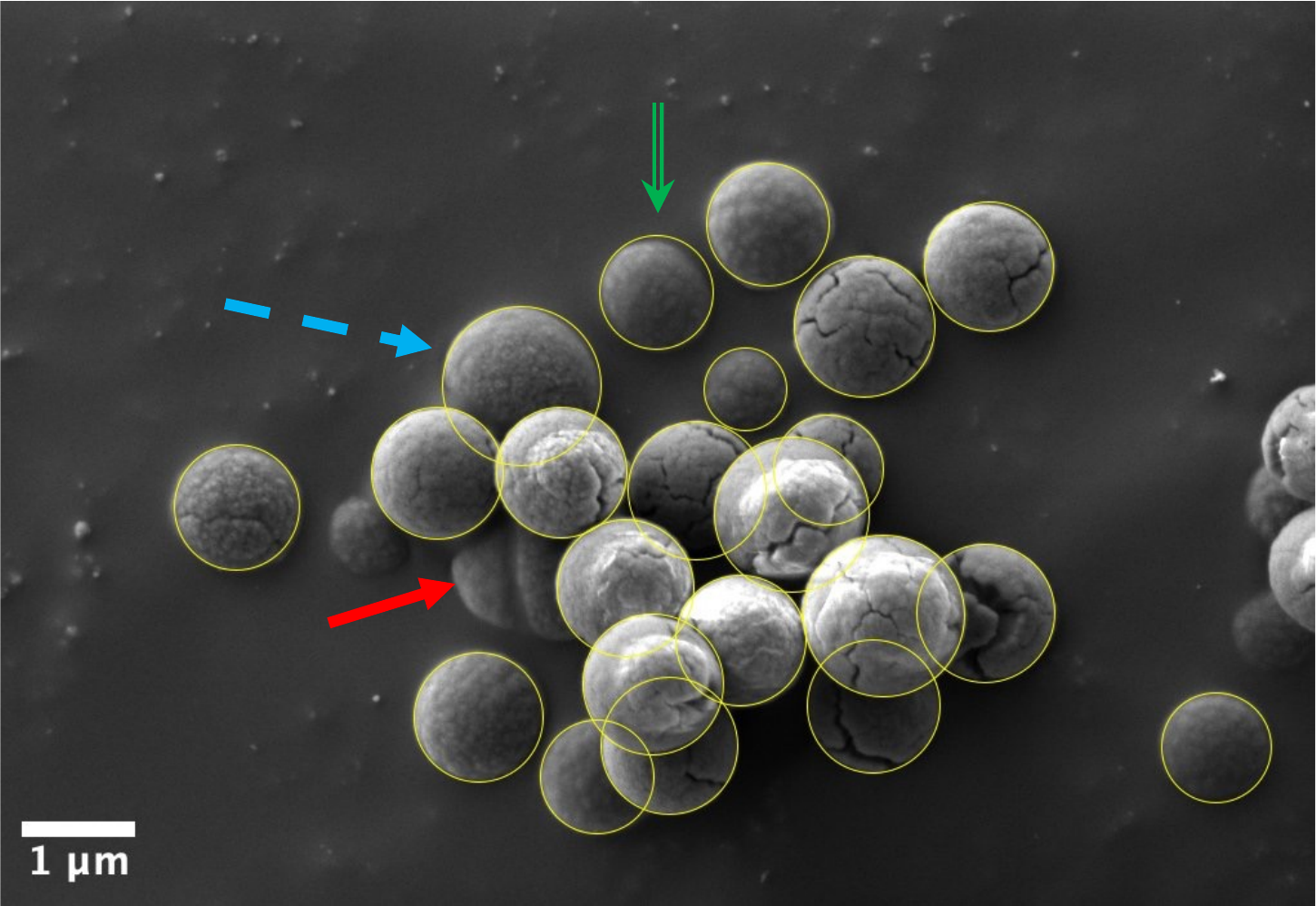}{0.582\textwidth}{(a)}
\fig{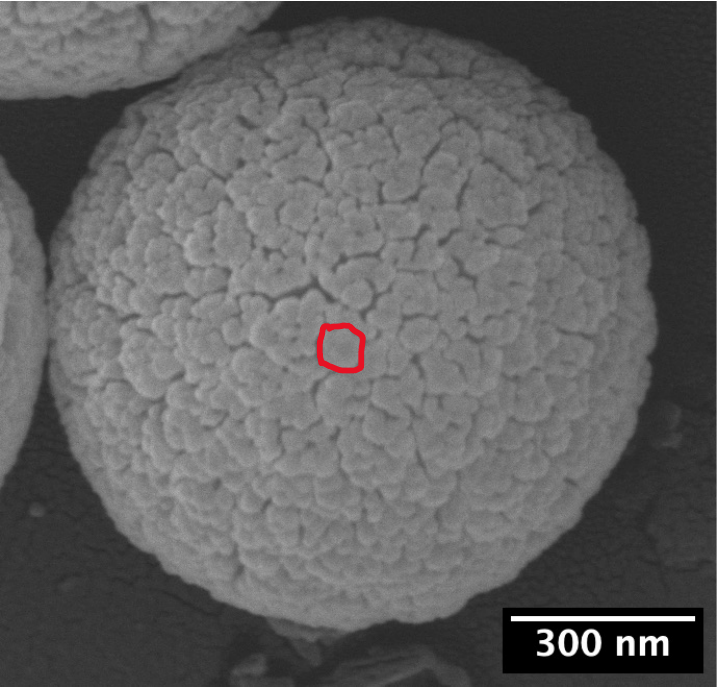}{0.418\textwidth}{(b)}}
\gridline{
\fig{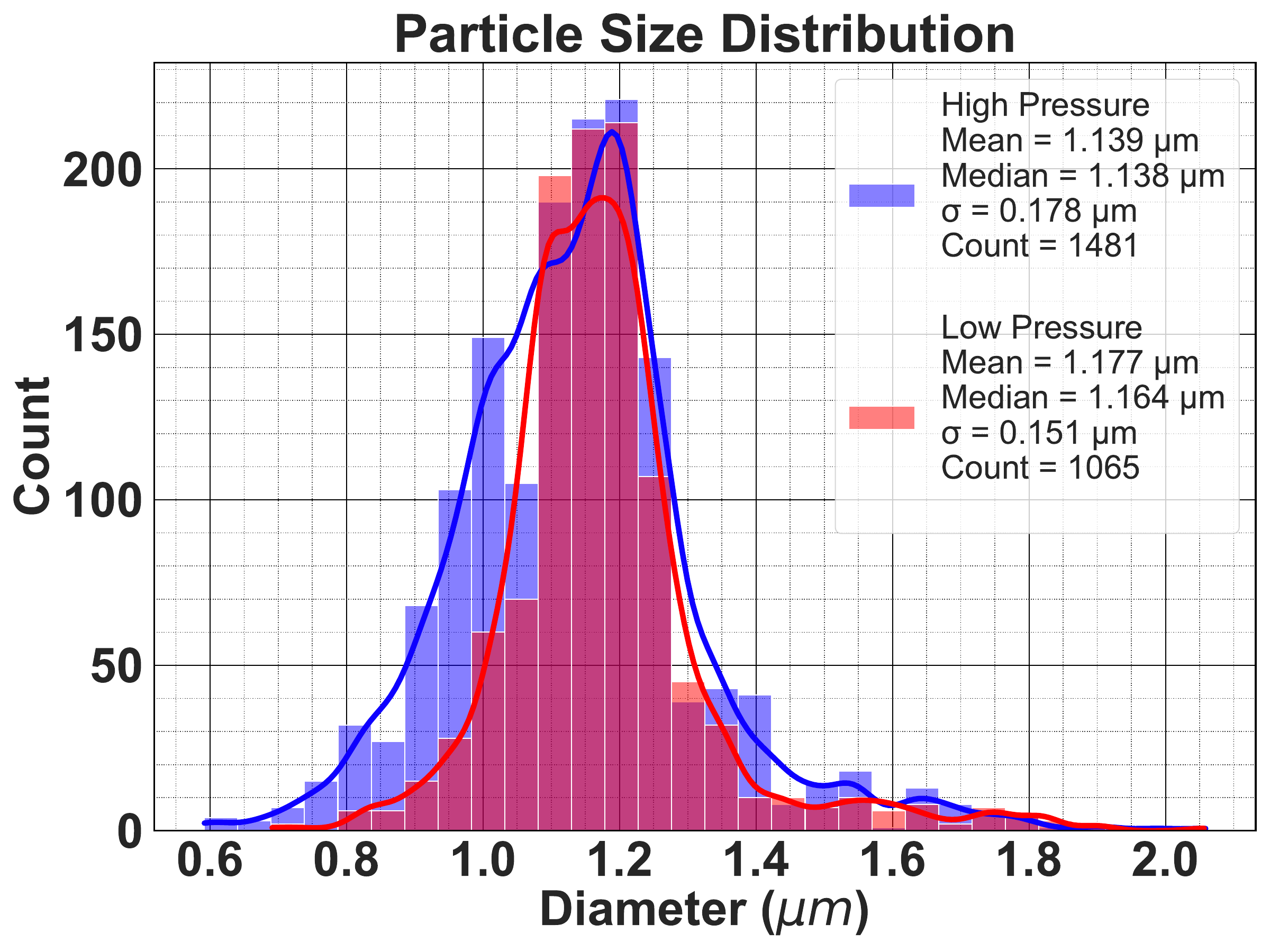}{0.75\textwidth}{(c)}}
\caption{\textbf{(a)} An example of a measured image for particle sizes, with corrected brightness and contrast. The yellow circles indicate a measured particle. The solid red arrow indicates an example of a particle that is too malformed to be counted. The dashed blue arrow indicates a particle that is obscured but can be measured by fitting the circle perimeter to the visible particle edges (<2\% of particles counted). The hollow green arrow indicates a particle that is not perfectly spherical, but has a counted diameter using the equivalent circular diameter (<1\% of particles counted). \textbf{(b)} A gold-coated tholin particle with sufficient resolution to image and count the tholin monomers, one of which is outlined in red. The monomers are the smaller units with clear separations that constitute the larger, spherical tholin particle. \textbf{(c)} Particle size distributions for high-pressure (blue) and low-pressure (red) aggregates. 1481 high-pressure and 1065 low-pressure particles were counted to create statistically significant sample sizes. The overlaid curves are the kernel density estimations of the histograms, using a Gaussian kernel.}
\label{fig:particlesizes}
\end{figure}

The particle size distributions of the low- and high-pressure samples are shown in Figure \ref{fig:particlesizes}c. The mean particle sizes, $1.139$~$\upmu$m and $1.177$~$\upmu$m for high- and low-pressure samples, respectively, are remarkably similar and are within a 1$\sigma$ standard error of each other. Across both samples, the size range is 0.59~$\upmu$m to 2.06~$\upmu$m. These sizes are in line with previously published tholin particle sizes  \citep[e.g.,][]{BarNun1988, Scattergood1992, ClarkeFerris1997, Szopa2006, Sciamma2017, Horst2018}. Additionally, some particles exhibit fracture features of various sizes. Some previously published works \citep[e.g.,][]{Vanssay1999} also show tholin particles with these features, while others \citep[e.g.,][]{Carrasco2009} do not. The source of this fracture may be due to the pressure-cycling that samples experience as they are taken from atmospheric pressure to vacuum, which would then cause any gasses trapped during formation to burst the particle during escape.

The monomer sizes are difficult to measure (see an example in Figure~\ref{fig:particlesizes}b) and therefore the total number of monomers measured are relatively low in number (${\sim}$30), but range from 20 to 150~nm. These sizes align with values seen both in the lab \citep{Trainer2006}, and with sizes from \textit{in-situ} Huygens Descent Imager/Spectral Radiometer (DISR) measurements \citep{Tomasko2005, Tomasko2008} and haze optical property modeling based on those DISR measurements \citep{Lavvas2010}.

\subsection{Density}\label{sec:densityresult}

\begin{figure}
\gridline{
\fig{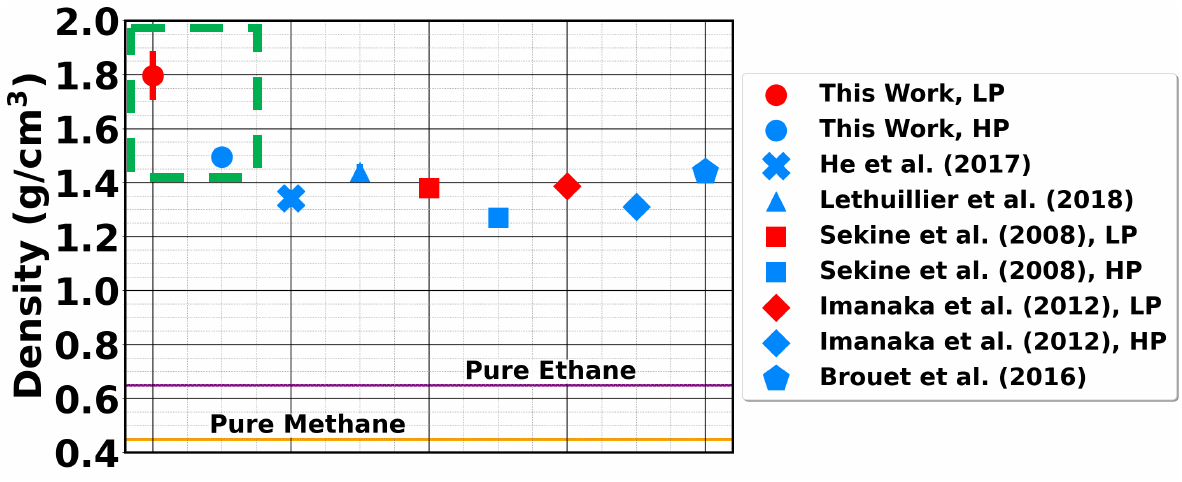}{0.7\textwidth}{(a)}}
\caption{Summary of helium pycnometry density data for tholin samples created using the 95\% \ce{N2}/5\% \ce{CH4} initial gas mixture. The green bounding box indicates data obtained for this study. Previously published pycnometry data for tholin samples produced with the same gas mixture are provided for comparison \citep{Sekine2008, Imanaka2012a, Brouet2016, He2017, Lethuillier2018}. Red points indicate lower-pressure (LP) samples and blue points indicate higher-pressure (HP) samples. The purple line indicates the density of pure liquid ethane between 90 and 95~K ($\rho=0.646-0.652$~g$\cdot$cm$^{-3}$). The orange line indicates the density of pure liquid methane between 90 and 95~K ($\rho=0.445 - 0.452$~g$\cdot$cm$^{-3}$). Pure liquid methane and ethane densities were calculated using Tables 2 and 8 from \citet{Yu2023}.}
\label{fig:density}
\end{figure}

We present density results for both low- and high-pressure samples in Figure \ref{fig:density} and Table \ref{tab:DensityTable}. Our density results are $\rho=1.496$ and $1.797$~$\text{ g$\cdot$cm}^{-3}$ for high- and low-pressure samples, respectively; In other words, our high-pressure sample has a lower density than our low-pressure sample. Additionally, these measurements are higher than previously published pycnometry measurements for tholins created with the same 95\% \ce{N2}/5\% \ce{CH4} initial gas mixture in other laboratories \citep{Sekine2008,Imanaka2012a,Brouet2016,He2017,Lethuillier2018}. Our density results are similar to those of organic materials on Earth, such as melamine ($\rho=1.593$~$\text{ g$\cdot$cm}^{-3}$, \citet{Haynes2017}). Additionally, all reported tholin density values thus far are within the typical organic molecule density range of 0.5 to 2.5~g$\cdot$cm$^{-3}$. These results suggest that if organic materials on Titan have a similar density, they would sink in Titan's ethane-methane-nitrogen lakes. However, this only considers the particle's density (i.e., Archimedes' buoyancy) and does not account for other factors, such as capillary forces or buoyancy effects from voids or gas pockets \citep{Cordier2019}.

\subsection{Surface Free Energy}\label{sec:SurfaceEnergy}

\begin{figure}
\gridline{
\fig{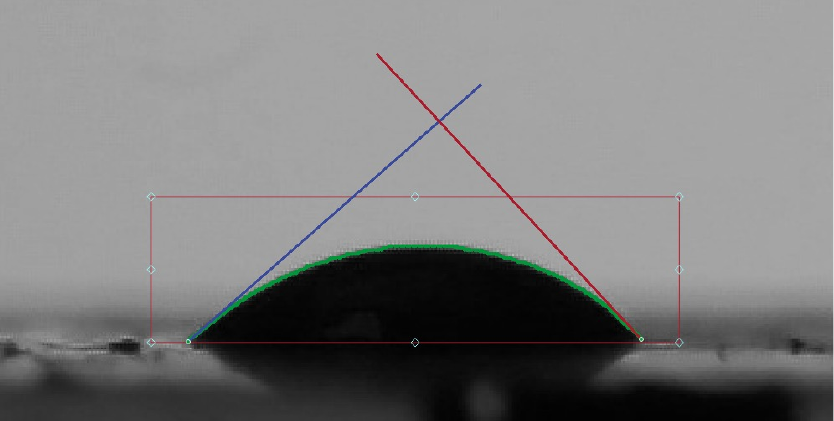}{0.6\textwidth}{(a)}}
\gridline{
\fig{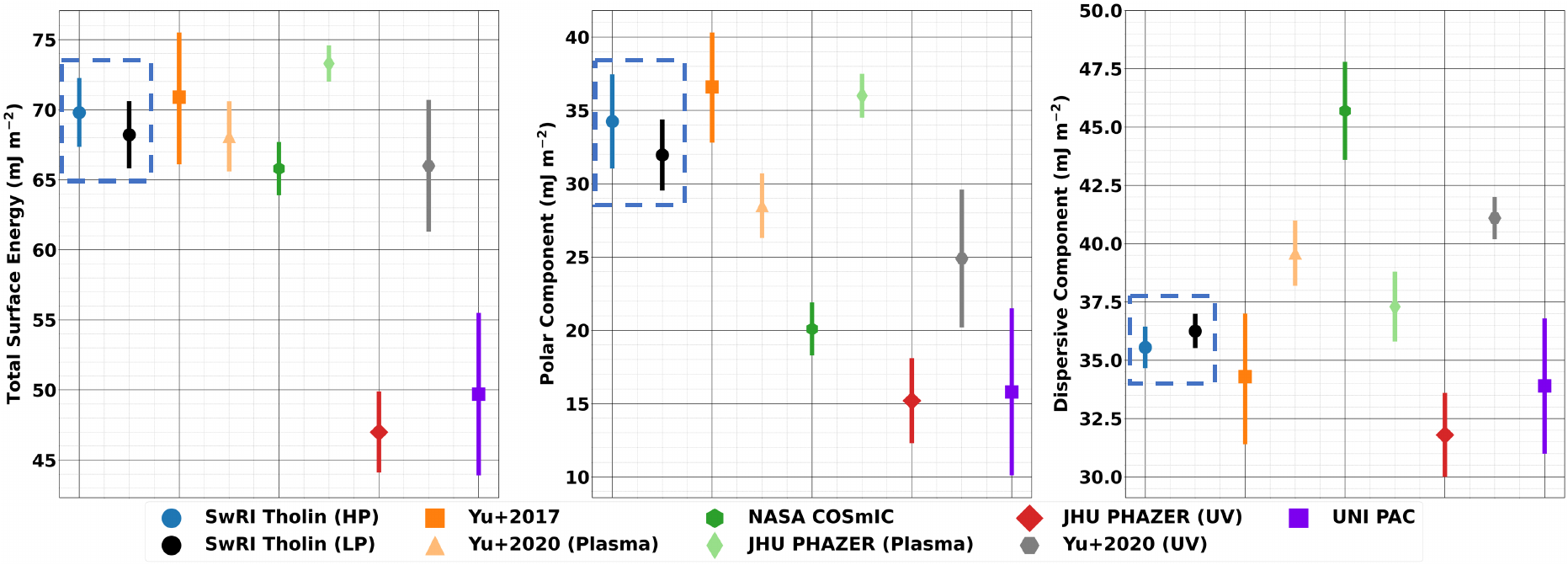}{1\textwidth}{(b)}
}
\caption{\textbf{(a)} An example of a one frame of a sessile droplet of diiodomethane and the contact angle it makes with a tholin film, as analyzed by Ossila Contact Angle Software 4.0. The red bounding box indicates the droplet area analyzed, with area outside of this bounding box not considered for the contact angle determination. The green curve indicates the outline of the analyzed droplet, and the black fill under the green curve indicates the detected droplet. The blue line indicates the left contact angle while the red line indicates the right contact angle. These angles are measured from the bottom of the bounding box to the respective left/right line.} \textbf{(b)} Summary of the total surface energy (left), polar component of the surface energy (middle), and dispersive component of the surface energy (right) of tholin samples. The red bounding box denotes the data obtained for this study. Total surface energy is given by summing the polar and dispersive components received from applying the OWRK method. Error bars for the samples used in this work are 1-$\sigma$ errors. Also included for comparison are surface energy results from the 5\%/95\% \ce{CH4}/\ce{N2} gas mixture tholins from \citet{Li2022}, for the PHAZER \citep[][${\sim}$2~Torr]{He2017}, NASA THS \citep[][${\sim}$22~Torr]{Sciamma-O'Brien2014}, and PAC \citep[][, ${\sim}$500~Torr]{SebreeC2018} experimental setups.
\label{fig:SurfaceEnergies}
\end{figure}

The surface energy values are provided in Figure \ref{fig:SurfaceEnergies}b. For comparison, we also plot values of the surface energies from \citet{Li2022}, which measured surface energy for the Johns Hopkins University (JHU)'s Planetary Haze Research (PHAZER) \citep{He2017}, NASA COSmIC/Titan Haze Simulation (THS) \citep{Sciamma2010}, and University of Northern Iowa (UNI)'s Planetary Aerosol Chamber (PAC) \citep{SebreeC2018} experimental setups. The OWRK method produces a result for the dispersive and polar components of the surface energy, which may then be summed to provide the total surface energy. 

For the high-pressure tholin sample, the individual dispersive ($\gamma_d)$ and polar components ($\gamma_p$) are $35.55$ and $34.25$~$\text{mJ$\cdot$m}^{-2}$, respectively. For the low-pressure sample, the individual dispersive and polar components are $36.25$ and $31.96$~$\text{mJ$\cdot$m}^{-2}$, respectively. The total surface energy ($\gamma_T = \gamma_d + \gamma_p$) are $69.80$ and $68.22$~$\text{mJ$\cdot$m}^{-2}$, for high- and low-pressure samples respectively. The measured surface energies and partitioning components between high- and low-pressure samples are within the standard errors of each other.

In general, we find agreement in the polar, dispersive, and total surface energies for the experimental setups that also use cold RF plasma as the energy source. NASA COSmIC/THS plasma samples have similar total surface energy as other plasma samples, but have a higher dispersive component and a lower polar component. This discrepancy is thought to be due to a significantly lower residence time of gas within the plasma for the NASA COSmIC/THS experiment, which leads to a different overall tholin chemistry, compared to the others \citep{Sciamma-O'Brien2014}.

\subsection{Young's Modulus and Nanoindentation Hardness}\label{sec:EandHResults}

\begin{figure}
\centering
\gridline{
\fig{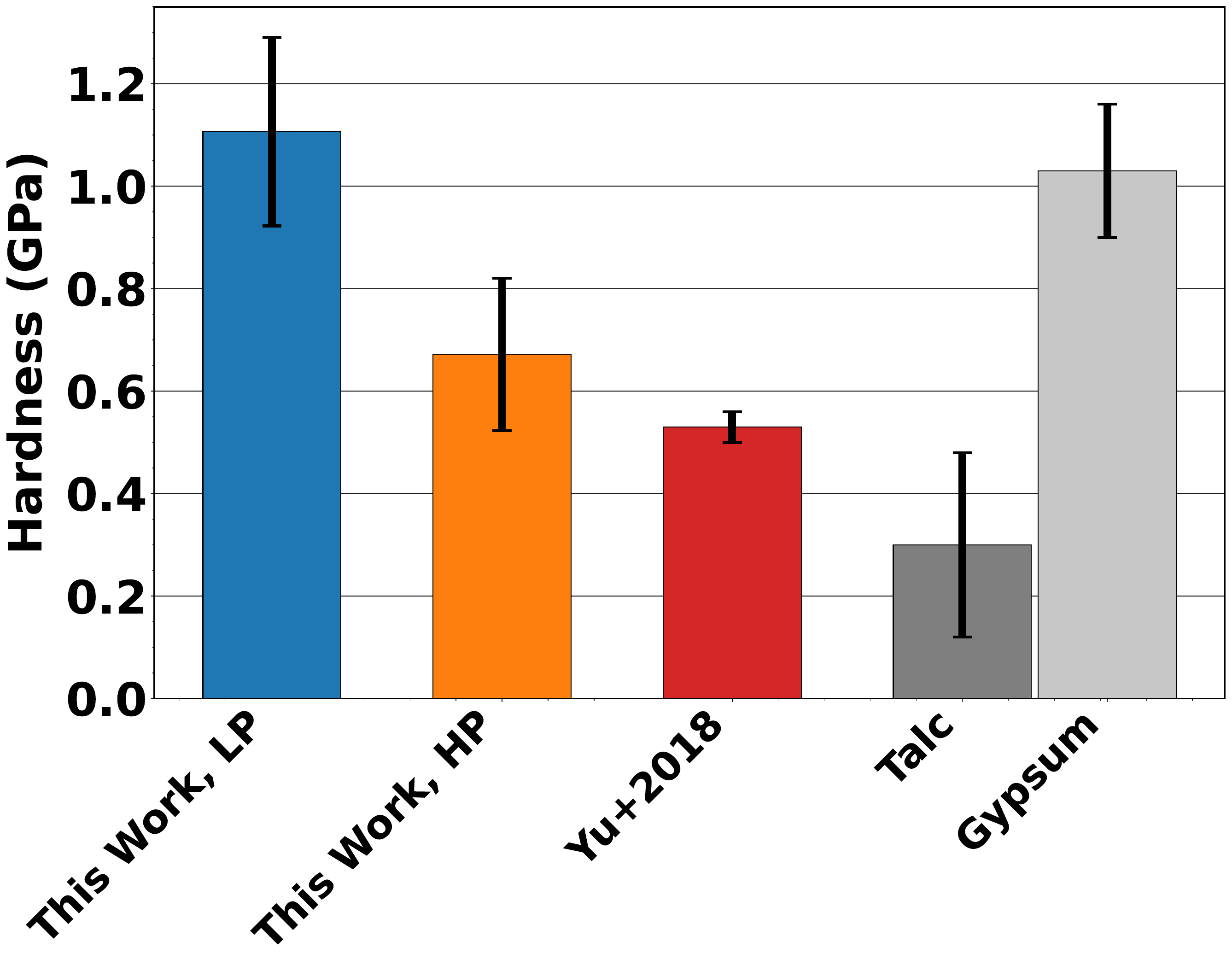}{0.45\textwidth}{(a)}
\fig{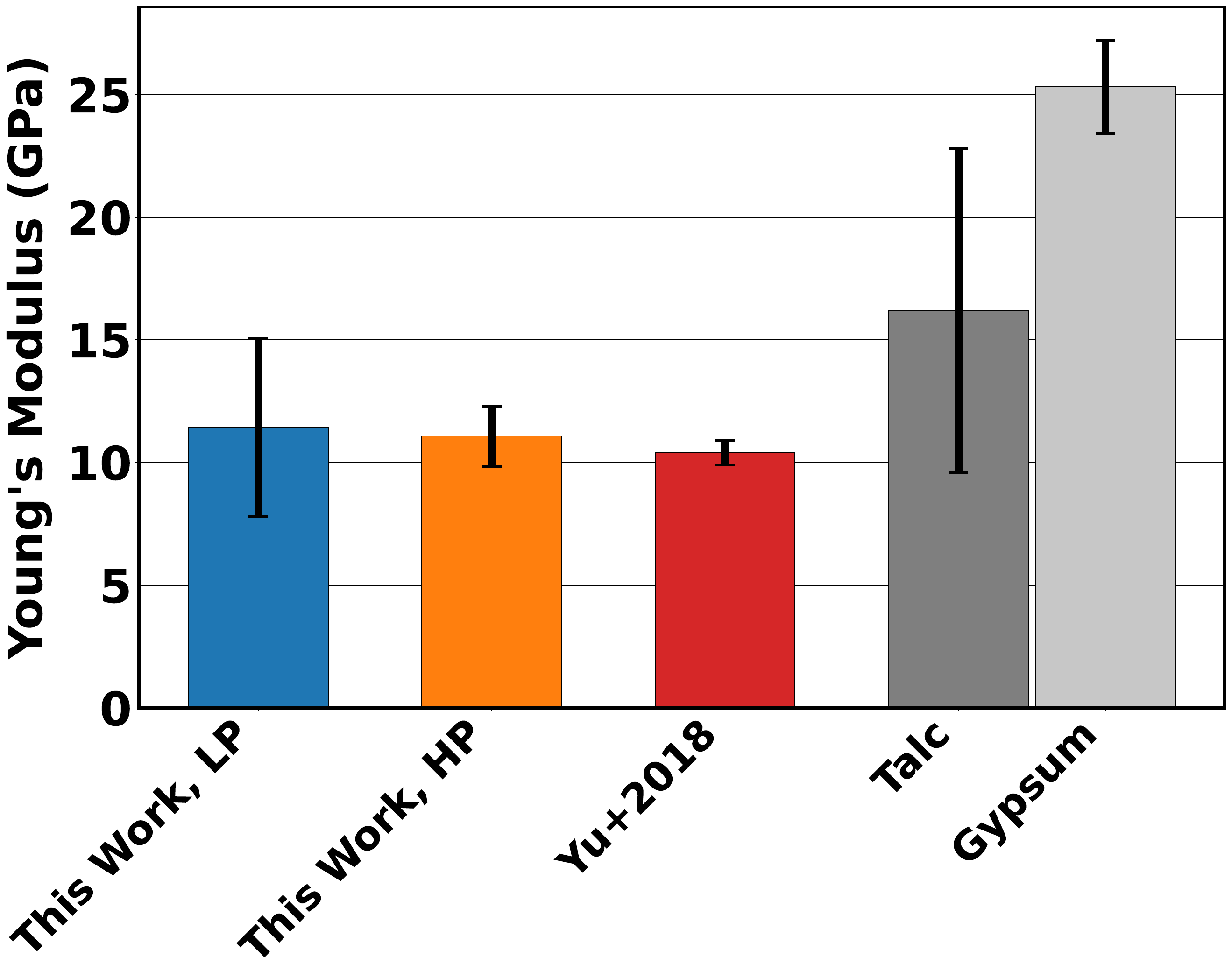}{0.45\textwidth}{(b)}
} 
\caption{\textbf{(a)} Nanoindentation hardness and \textbf{(b)} Young's modulus results from dynamic nanoindentation of the tholin film samples produced in this work. Error bars are 1-$\sigma$ errors. For comparison is the tholin sample produced and measured by \citet{Yu2018}, as well as Talc and Gypsum nanoindentation data from \citet{Broz2006}.}
\label{fig:HardnessModulus}
\end{figure}

Figure \ref{fig:HardnessModulus} summarizes the Young's modulus and nanoindentation hardness results of both tholin sample sets synthesized for this study, as well as tholin samples previously measured in \citet{Yu2018}. Additionally, relevant Earth or organic materials such as silicate sand, gypsum sand, and melamine \citep{Yu2018}, and talc and gypsum minerals \citep{Broz2006} are included for comparison. 

For high-pressure samples, we obtained a Young's modulus and nanoindentation hardness of $E_{HP}=11.07$~GPa and $H_{HP}=0.67$~GPa, respectively. For low-pressure samples, we obtained a Young's modulus and nanoindentation hardness of  $E_{LP}=11.43$~GPa and $H_{LP}=1.11$~GPa, respectively. We see that the high-pressure films are mechanically weaker than the low-pressure films, with the low-pressure films being 65\% harder and having a 3\% higher Young's modulus than the high-pressure films.

In general, we see a continued trend of low resistance to applied stress, abrasion, and deformation of tholin samples, indicated by their lower Young's modulus and nanoindentation hardness compared to common Earth sediments. Our tholin samples have similar mechanical properties to the previously published tholin results, with $E{\sim}10$~GPa and $H{\sim}0.5$~GPa \citep[][]{Yu2018}. Both samples are at or below the Young's modulus and nanoindentation hardness of some of the weakest minerals on Earth, including talc \citep[$E{\sim}16.2$~GPa, $H{\sim}0.3$~GPa,][]{Broz2006} and gypsum \citep[$E{\sim}25.3$~GPa, $H{\sim}1.03$~GPa,][]{Broz2006}.  Silicate sand, a major component of most Earth deserts and dunes, has a significantly higher modulus and nanoindentation hardness \citep[$E{\sim}105.60$~GPa, $H{\sim}13.539$~GPa,][]{Yu2018} than any tholin samples measured thus far. 

\section{Discussion}\label{sec:discussion}
\subsection{High- and Low-Pressure Sample Differences}\label{sec:hplpdiff}
To investigate the differences between high- and low-pressure samples, we calculate a two-tailed Z-score using
\begin{equation}
    z=\frac{\left|\mu_{HP}-\mu_{LP}\right|}{\sqrt{\frac{\sigma_{HP}^2}{n_{HP}}+\frac{\sigma_{LP}^2}{n_{LP}}}}
\end{equation}
where $\mu$, $\sigma$, and $n$ are the mean, standard deviation, and number of samples respectively. Values of $z\geq2$ correspond to a two-tailed p-value of $p\approx0.045$ and are therefore considered significant. In summary, both the high-pressure and low-pressure samples exhibit practically similar particle sizes and morphologies ($1.139\pm0.178$~$\upmu$m vs $1.177\pm0.151$~$\upmu$m, $z=0.2$, $p=0.84$), surface free energies ($69.80\pm2.45$~mJ$\cdot$m$^{-2}$ vs $68.22\pm2.4$~mJ$\cdot$m$^{-2}$, $z=1.0$, $p=0.32$), and Young's modulus ($11.07\pm0.61$~GPa vs $11.43\pm1.81$~GPa, $z=0.2$, $p=0.84$). In contrast, the samples differ in production rate (6~mg$\cdot$hr$^{-1}$ vs 2~mg$\cdot$hr$^{-1}$), density ($1.797\pm 0.091$~g$\cdot$cm$^3$ vs $1.496\pm0.005$~g$\cdot$cm$^3$, $z=3.3$, $p\approx0.001$), and nanoindentation hardness ($0.67\pm0.07$~GPa vs $1.11\pm0.09$~GPa, $z=3.9$, $p<0.001$). More simply, the low-pressure sample has a lower production rate, higher density, and higher mechanical strength. We are interested in understanding what causes the similarities and differences between the high- and low- pressure samples. Furthermore, we want to explore the fundamental cause for the differences between the samples, such as their chemical compositions, cross-linking ratios, and porosities.

Variations in experimental pressure can lead to changes in gas density, collisional frequency, plasma density and uniformity, and gas residence time. The fact that both the high- and low-pressure samples have a mean size of around 1.1~$\upmu$m indicates that residence time likely has minimal effect on the resulting particle properties. As mentioned in Section \ref{sec:synthesis}, the residence time may be estimated using Equation \ref{eq:residencetime}. The Tholinator has a gas residence time of approximately 47~s at 1~Torr and 16~s at 0.125~Torr. Despite this threefold reduction in residence time and tenfold reduction in pressure, the particle size distributions remain very similar across both samples. This may indicate that particle sizes are a result of chemical processes or other experimental factors rather than residence time alone. For comparison, the COSmIC/THS experimental setup, which has a residence time of ${\sim}3$~$\upmu$s, generates particles in the range of 100 to 500~nm \citep{Sciamma2017}. This residence time is so short that only the initial stages of the aerosol chemistry occur, limiting the resulting chemistry \citep{Sciamma-O'Brien2014, Sciamma2017}. Regardless of these differences, when compared with the Tholinator which has a residence time $5$-to-$14\times10^6$ times longer, the particle sizes roughly double from 500~nm to $1$~$\upmu$m, showing a weak link between residence time and particle size. 

Since many parameters covary with pressure, we cannot tell whether gas density, collisional frequency, and/or plasma density lead to the observed changes between the high- and low-pressure samples. Several recent works have begun to understand some of these other parameters and their effects on resulting composition, morphology, and particle sizes \citep[][]{He2022, Perrin2025}.

The similarity in size and morphology between the two sample sets (Section \ref{sec:psd}), despite their differing densities (Section \ref{sec:densityresult}), suggests that the low-pressure samples may exhibit either a compositional difference, lower porosity, and/or higher cross-linking rate compared to the high-pressure samples. Previous works have shown that the high- and low-pressure samples likely have compositional differences, as both the gas-phase and solid phase products show differences in C/N and C/H ratios \citep{Imanaka2004, Sekine2008, He2022}. These different C/N and C/H ratios could lead to tholin samples with altered mechanical strength and density. 

However, because the two samples have different densities and mechanical strengths, we suspect cross-linking and porosity may also be different between the samples. Previous works studying plasma polymerization have shown that plasma discharge can lead to cross-linking of the samples \citep[e.g.,][]{Hudis1972, Friedrich2011}. Cross-linking does not necessarily alter the bulk morphology of polymers but would create a denser material and reduce the amount of voids present \citep[e.g.,][]{Friedrich2011, Winning2024}, which is consistent with the similar particle size and different densities of our low- and high-pressure samples. Additionally, cross-linking also has effects on Young's modulus and nanoindentation hardness, typically increasing their values for more cross-linked polymers \citep[e.g.,][]{Nielsen1969, Aryaei2012, Wang2014, Chen2023}, while less porous material typically exhibits higher Young's modulus and nanoindentation hardness \citep[e.g.,][]{Roberts1991, JangMatsubara2005, Zhang2014, Balak2015}. This behavior aligns with the observed properties of our samples, leading us to believe that the low-pressure sample may be less porous and more cross-linked than the high-pressure sample.

Unfortunately, the parameter space to explore is large, but future work should focus on investigating the chemical differences, true (rather than skeletal) density, porosity, and cross-linking percentage of the high- and low-pressure samples. Additionally, more work should be done to explore the wide range of experimental parameters, such as collisional frequency and plasma density, to understand the full connection between experimental parameters to the resulting sample properties. 
\subsection{Implications for Organics on Titan}\label{sec:implication}
From a holistic perspective, we may make a few predictions about organics on Titan, with the assumption that they are similar to the tholin samples created in the lab. If they are of a similar size to tholins and atmospheric haze organics, the surface organics on Titan would need to undergo some process of particle size growth and hardening in order to form the equatorial dunes observed by Cassini. These dune particles are estimated to have an ideal saltation size (i.e., the size where the required wind speed to saltate is minimized) of approximately 100 to 1000~$\upmu$m \citep{Lorenz2006, Lorenz2013, Burr2015}, while lab organics are only ${\sim}1$~$\upmu$m in size, with our maximum size being 2.06~$\upmu$m.

Because a portion of Titan's hazes form in the upper atmosphere at much lower pressures (as low as ${{\sim}}10^{-6}$~Torr, or ${\sim}10^{-4}$~Pa), the actual haze particles may possess even higher density and mechanical strength than our laboratory analogs, with important consequences for various atmospheric and surface processes on Titan. Previous work has demonstrated that haze particle floatation in Titan's lakes is unlikely due to the positive density contrast between the haze particles and the lake liquids (liquid methane/ethane/nitrogen mixture) \citep{Li2022,Yu2020}. A positive capillary force is thus needed to support floatation, which in turn demands a high liquid-solid contact angle \citep{Cordier2019}. The actual haze particles, which are formed at even lower pressures than our current experimental conditions, may have even higher density, producing an even larger density contrast with the lakes, thus requiring an unrealistically high contact angle for floatation. Our surface energy measurements indicate near-complete wetting (contact angle ${\sim}$0$^\circ$), confirming that floatation is unlikely. In addition, our experimental results support that the haze particles seem to become less porous when produced at lower pressure, suggesting that the actual Titan haze particles, which are formed at orders of magnitude lower pressures, are unlikely to have sufficient porosity to allow for gas-filled pores to provide buoyancy as suggested by \citet{Yu2024}. Thus, we believe the haze particles would likely sink into the lakes.

Our results confirm previously published results \citep{Yu2018} that show that Titan's aerosol analogs are mechanically weaker than minerals on Earth which typically form sands and dunes. As a comparison, gypsum sand \citep[$E{\sim}37.44$~GPa, $H{\sim}1.531$~GPa,][]{Yu2018}, a less common form of sand that forms dunes in places such as White Sands National Park in New Mexico and Guadalupe Mountains National Park in Texas, is weaker and more flexible than silicate sand \citep[$E{\sim}105.60$~GPa, $H{\sim}13.539$~GPa,][]{Yu2018} but still has a Young's modulus and nanoindentation hardness that are a factor of two greater than the stiffest and hardest tholin samples. The low Young's modulus and nanoindentation hardness may mean that, over time, organics on Titan could form very fine particulates on Titan's surface. Without additional mechanisms to harden the samples or increase their size, the small sizes (${\sim}1$~$\text{$\upmu $m}$) and low abrasion resistance suggest that unaltered organic aerosols on Titan, if they have similar values, would not be sufficient to form Titan's equatorial dunes \citep[e.g.,][]{MacKenzie2023}. Previous work has suggested mechanisms that may lead to dune formation using Titan's aerosols, such as sintering \citep[][]{Barnes2015, Lapotre2022} and flocculation \citep[][]{Barnes2015, Hirai2023}.

However, our results also show that our samples produced at pressures of 0.125~Torr (16.66~Pa) are mechanically stronger than those produced at 1~Torr (133.3~Pa), and Titan's hazes form under much lower pressures still (${\sim}10^{-6}$~Torr/${\sim}10^{-4}$~Pa). This trend might imply that actual Titan haze particles are harder and stiffer than our laboratory analogs. If that is the case, it may be possible that these lower-pressure aerosols contribute to an overall material that is much harder once settled onto Titan's surface. However, this inference assumes that pressure differences are the dominant or sole factor influencing haze mechanical properties. 

In reality, other variables such as total irradiation, electron density, and ionization efficiency also vary with pressure due to atmospheric absorption effects \citep[e.g.,][and references therein]{Coates2009,Krasnopolsky2014,MitchellLora2016}. This changing environment could lead to enhancement or depletion of certain molecules, potentially leading to altered end product aerosols. Previous work that investigated tholins under different enrichment or starvation conditions have shown that even small differences of certain molecules, such as \ce{CO} or \ce{C2H2}, lead to particles with different chemical composition, densities, or overall particle sizes \citep[e.g.,][and references within]{He2017,Ugelow2024,Dubois2025}.

Critically, our dataset includes only two pressure conditions, which limits our ability to capture the full relationship between formation pressure and the physical properties measured. It remains unclear whether the observed trend persists toward lower pressures, reverses, or plateaus beyond the range explored here. Future work should aim to explore the properties of tholins formed under even lower pressures, although this may require unrealistic runtimes or even fail to produce solid material.

Finally, our results have broader implications for other haze-forming bodies in the Solar System and exoplanets that have N$_2$/CH$_4$ atmospheres. As an example, Triton and Pluto both have a similar atmospheric composition to Titan, with a major N$_2$ component and minor CH$_4$ and CO components \citep[e.g.,][]{Broadfoot1989, Gladstone2019}. Similarly to Titan, the \ce{N2}/\ce{CH4} atmosphere leads to the formation of organic aerosols in the atmosphere \citep{Hillier1991, Cheng2017}, which may then settle onto the surface \citep[e.g.,][]{Grundy2018}. However, unlike Titan's 1.5~bar atmosphere, Triton and Pluto have tenuous atmospheres, with surface pressures of $14$~$\upmu$bar and $11$~$\upmu$bar respectively \citep{Gladstone2016,Sicardy2024}. Therefore, low-pressure aerosols are likely the dominant form of haze on Triton and Pluto, whereas Titan hosts a broader range of aerosols formed under high- and low-pressure regimes spanning many magnitudes of pressure.

\section{Conclusions}

The Cassini--Huygens mission revealed Titan to be one of the most interesting moons across multiple disciplines, detecting ions as large as $10,000$~amu/q at pressures as low as $10^{-6}$~Torr, an Earth-like geology, a liquid methane-ethane cycle similar to Earth's hydrological cycle, and a smothering atmosphere with ubiquitous haze layers. Titan's hazes and their potential interactions with Titan's surface features have been the subject of many studies after Cassini. In an effort to understand Titan's hazes, analogs known as `tholins' have been in production since the 1970s. Here we have introduced one such analog-producing experimental system, known as the Tholinator, at the Southwest Research Institute's Space Environment Simulation Laboratory. Previously established tholin generation experiments employed pressures much higher than Titan's aerosol-forming regions, at ${\sim}1$~Torr or higher, which thus far had an unknown effect on the resulting aerosols'  physical properties. Here, we produced tholins at two pressures, 1~Torr and 0.125~Torr, and measured their production rate, particle size, morphology, surface free energy, density, Young's modulus, and nanoindentation hardness, to quantify the effects of the formation pressure on these physical properties of tholins. We found that:

\begin{enumerate}
    \item Solid powder production rates of this new experimental setup are roughly similar to those of other plasma-based tholin production experiments, producing 6~mg$\cdot$hr$^{-1}$ at 1~Torr and 2~mg$\cdot$hr$^{-1}$ at 0.125~Torr.
    \item Both high- and low-pressure samples have an average particle size of $1.1$~$\upmu$m. Previously published setups have similar particle sizes and morphologies, including a fracture feature that may be due to vacuum pressure cycling.
    \item Low-pressure and high-pressure samples have similar total surface free energies ($\gamma{\sim}69$~mJ$\cdot$m$^{-2}$) and comparable polar (${\sim}33$~mJ$\cdot$m$^{-2}$) and dispersive (${\sim}36$~mJ$\cdot$m$^{-2}$) components. Previously published samples have similar total surface free energies but slightly differing polar and dispersive components.
    \item Low-pressure samples have a higher density ($1.797$~$\text{g$\cdot$cm}^{-3}$) than high-pressure samples ($1.496$~$\text{g$\cdot$cm}^{-3})$. This density difference has been observed previously by \citet{Imanaka2004} and \citet{Sekine2008}. 
    \item The low- and high-pressure samples have similar Young's moduli ($E_{LP}=11.43$~GPa vs $E_{HP}=11.07$~GPa). However, the low-pressure tholin sample has a higher nanoindentation hardness ($H_{LP}=1.11$~GPa) than the high-pressure tholin sample ($H_{HP}=0.67$~GPa).
\end{enumerate}

In general, we find this new experimental setup generates materials with properties that agree with previously published results in particle size, particle morphology, surface energy, and Young's modulus, especially when limiting comparisons to tholins created under similar conditions, i.e., similar pressures, gas mixtures, and energy sources. Our high-pressure sample differs from the low-pressure sample in density and nanoindentation hardness, which may be due to a difference in sample porosity, cross-linking, C/H ratio, and C/N ratio. Future work should focus on uncovering the mechanisms responsible for these differences, quantifying the relative contribution of each factor, and investigating how these effects evolve with formation gas pressure. In particular, it will be important to determine whether the trend toward harder and less porous aerosols continues at lower pressures.

\section{Data Availability}

Data is already available through the Zenodo repository (\url{https://doi.org/10.5281/zenodo.18227163}) and appendix data tables. Additional data may be supplied upon reasonable request.

\section{Acknowledgements}
X. Yu and A. Husi\'c are supported by the NASA Cassini Data Analysis Program Grants 80NSSC21K0528 and 80NSSC24K0203. X. Yu is also supported by the NASA Planetary Science Early Career Award 80NSSC23K1108, the NASA Solar System Workings Grant 80NSSC24K0888, the NASA Planetary Data Archiving, Restoration and Tools Grant 80NSSC25M7002, the NASA Emerging Worlds Grant 80NSSC26K0080, and the Heising-Simons Foundation grant 2023-3936. We would like to thank Brenna Halverson and Austin Patridge for their support in performing helium pycnometry measurements, and Ana Stevanovic of the Kleberg Advanced Microscopy Center for help with scanning electron microscopy imaging. This work also benefited from the Texas Area Planetary Science Meeting (TAPS) Series hosted by the University of Texas at San Antonio, a conference series funded by the Heising-Simons Foundation. We thank our anonymous reviewers for their feedback which resulted in a better final manuscript.

\bibliography{bibliography}{}
\bibliographystyle{aasjournal}

\appendix
\section{Data Tables}
Below are the data tables used to create the resultant figures in Section \ref{sec:results}.

\begin{deluxetable*}{lcc}[htb]
\tablecaption{Density results for SwRI Tholinator tholin samples, compared to previously published 95\% N$_2$/5\%CH$_4$ plasma samples. Pressure refers to tholin synthesis gas pressures, if available. Errors for SwRI samples are reported as 1-$\sigma$ errors.
\label{tab:DensityTable}}
\tablehead{\colhead{Sample} & \colhead{Pressure} & \colhead{Density} \\ 
\colhead{} & \colhead{(mTorr)} & \colhead{(g/cm$^3$)} } 
\startdata
SwRI 022224 & 960 & 1.525 $\pm$ 0.066\\
SwRI 150424 &  960 & 1.716 $\pm$ 0.123 \\
SwRI 052224A & 960 & 1.379 $\pm$ 0.248 \\
SwRI 052224B & 140 & 1.797 $\pm$ 0.091 \\
\citet{Sekine2008} &  300 &  1.38 \\
\citet{Sekine2008} &  1125 &  1.27 \\
\citet{He2017} &  2000 &  1.343 $\pm$ 0.002 \\
\citet{Imanaka2012a} &  195 &  1.3866 $\pm$ 0.0004 \\
\citet{Imanaka2012a} &  1200 &  1.3113 $\pm$ 0.0003  \\
\citet{Brouet2016} &  N/A &  1.440 $\pm$ 0.01 \\
\citet{Lethuillier2018} &  N/A &  1.440 $\pm$ 0.03 \\
\enddata
\end{deluxetable*}

\begin{deluxetable*}{lcc}[htb]
\tablecaption{Young's Modulus (E) and Nanoindentation Hardness (H) data for SwRI Tholinator tholin samples, compared to blank substrates, Earth sands, Talc, Gypsum, and the tholin sample from \citet{Yu2018}. Errors for SwRI samples are reported as 1-$\sigma$ errors.
\label{tab:EandHTable}}
\tablehead{\colhead{Sample} & \colhead{Young's Modulus, E} & \colhead{Nanoindentation Hardness, H} \\ 
\colhead{} & \colhead{(GPa)} & \colhead{(GPa)} } 
\startdata
SwRI High-Pressure Film & 11.07 $\pm$ 0.61 & 0.67 $\pm$ 0.07\\
SwRI Low-Pressure Film & 11.43 $\pm$ 1.81 & 1.11 $\pm$ 0.09\\
Si Wafer, blank\tablenotemark{a} & 141.06$\pm$ 9.56 & 12.78 $\pm$ 0.29\\
Epoxy Stub\tablenotemark{a} & 3.43 $\pm$ 0.13 & 0.19 $\pm$ 0.02 \\
Glass, blank\tablenotemark{a} & 71.45 $\pm$ 1.89 & 7.21 $\pm$ 0.15\\
Tholin\tablenotemark{b} & 10.4 $\pm$ 0.5 & 0.53 $\pm$ 0.03 \\
Talc\tablenotemark{c} & 16.2 $\pm$ 6.6 & 0.30 $\pm$ 0.18\\
Gypsum\tablenotemark{c}& 25.3 $\pm$ 1.9 & 1.03 $\pm$ 0.13\\
Melamine\tablenotemark{b}& $9.0\pm2.8$ & $0.48\pm0.21$\\
Silicate Sand\tablenotemark{b} & 105.60 $\pm$ 1.70& 13.539 $\pm$ 0.336  \\
Gypsum Sand\tablenotemark{c} & 37.44 $\pm$ 7.19  & 1.531 $\pm$ 0.402
\enddata
\tablenotetext{a}{Blank substrate materials from the same lot as substrates that have tholin films. For epoxy, this is the result of indenting areas of the stub that have no tholin material.}
\tablenotetext{b}{Data from \citet{Yu2018}.}
\tablenotetext{c}{Data from \citet{Broz2006}.} 
\end{deluxetable*}

\begin{deluxetable*}{lccc}[htb]
\tablecaption{Surface free energies of tholins produced for this study (SwRI Tholinator), as well as those from the JHU PHAZER experiment \citep{He2017}, NASA COSmIC/THS experiment \citep{Sciamma-O'Brien2014}, and UNI PAC experiment \citep{SebreeC2018}. Surface free energy data for non-SwRI samples are given by \citet{Yu2017, Yu2020}, and \citet{Li2022}. Errors for SwRI samples are reported as 1-$\sigma$ errors. 
\label{tab:SurfEnergyTable}}
\tablehead{\colhead{Sample} & \colhead{Total Surface Free Energy, $\gamma$} & \colhead{Dispersive Component, $\gamma_d$} & \colhead{Polar Component, $\gamma_p$}\\ 
\colhead{} & \colhead{(mJ$\cdot$m$^{-2}$)} & \colhead{(mJ$\cdot$m$^{-2}$)} & \colhead{(mJ$\cdot$m$^{-2}$)}} 
\startdata
SwRI, High Pressure & $69.80\pm0.99$ & $35.55\pm0.26$ & $34.25\pm1.00$\\
SwRI, Low Pressure & $68.22\pm0.88$ & $36.26\pm0.25$ & $31.96\pm0.89$\\
NASA COSmIC/THS & $65.8\pm1.9$ & $44.4\pm1.0$ & $20.1\pm1.8$ \\
JHU PHAZER, UV & $47.0\pm2.9$ & $31.8\pm1.8$ & $15.2\pm2.9$\\
JHU PHAZER, Plasma & $73.3\pm1.3$ & $37.3\pm1.5$ & $36.0\pm1.5$\\
UNI PAC & $49.7\pm5.8$ & $33.9\pm2.9$ & $15.8\pm5.7$\\
\citet{Yu2017} & $70.9^{+4.6}_{-4.8}$ & $34.3^{+2.7}_{-2.9}$ & $36.6^{+3.7}_{-3.8}$\\
\citet{Yu2020}, Plasma & $68.1\pm2.5$ & $39.6\pm1.4$ & $28.5\pm2.2$\\
\citet{Yu2020}, UV & $66.0\pm4.7$ & $41.1\pm0.9$ & $24.9\pm4.7$
\enddata
\end{deluxetable*}
\end{CJK*}
\end{document}